\documentclass[12pt,preprint]{aastex}

\newcommand{\xmm}{{\it XMM-Newton}}

\newcommand{\chandra}{{\it Chandra}}
\newcommand{\letg}{{\small LETGS}}

\newcommand{\agn}{{\small AGN}}

\newcommand{\hullac}{{\small HULLAC}}

\newcommand{\ebit}{{\small EBIT}}
\newcommand{\llnl}{{\small LLNL}}

\shorttitle{X-Ray Spectroscopy and Atomic Data}
\shortauthors{Behar \& Kahn}

\begin{document}

\title{X-Ray Spectroscopy and Atomic Data}


\author{Ehud Behar and Steven M. Kahn}
\affil{Columbia Astrophysics Laboratory, Columbia University, New York, NY}




\begin{abstract}
The Laboratory Astrophysics program employing the Lawrence Livermore National
Laboratory (\llnl) Electron Beam Ion Trap (\ebit) has been providing useful
atomic data in support of the x-ray missions \chandra\ and \xmm. 
Major achievements have been made for Fe-L ions in hot, collisional
plasmas, relevant to stellar coronae, supernova remnants, elliptical galaxies,
and galaxy clusters. 
Measurements for L-shell ions
of other cosmically important elements are also required, some of which are in
the \llnl\ \ebit\ pipeline. On the other hand, data for inner-shell excited
lines relevant to photoionized plasmas near accretion sources are largely
lacking. 
Even the wavelengths of these lines are only poorly known, which severely
limits their use for diagnostics, despite the great potential. 

\end{abstract}




\section{Introduction}

With the advent of the high-resolution grating spectrometers on board the
\chandra\ and \xmm\ x-ray observatories, spectroscopy has taken central
stage in the rejuvenated field of X-Ray Astronomy. The increasing role of
spectroscopy as a tool for astrophysical measurements has naturally drawn
attention to the relevant atomic physics and atomic data, as those are
directly linked with the ability to draw meaningful conclusions from observed
spectra. 
The x-ray wavelength band between 1 and 100 \AA, which is covered
complementally by the \chandra\ and \xmm\ spectrometers, contains a rich
forest of spectral lines emitted by highly charged ions that form at electron
temperatures of $kT_e$ = 0.1 - 3 keV. At these high temperatures, most of the
cosmically abundant elements from C to  Ni are stripped down to their K shell 
(i.e., $n$~= 1, $n$ being the principal quantum number) and emit relatively
few spectral lines. 
Additionally, many strong lines of L-shell ($n$~= 2) ions of
Si to Ni fall in this wavelength range. 
The eight ionization stages in the L shell can provide more precise
information on the temperature structure of the source, independent of
elemental abundances, than the two K-shell ionization stages are capable of.

\section{The Central Role of Fe-L}

The most prominent L-shell x-ray lines in astrophysical spectra are those of
iron. Fe-L line emission has been used recently to probe the hot temperature
structure in stellar coronal sources \citep {brinkman01}, supernova remnants 
\citep {kurt02} and galaxy clusters \citep {peterson01}. In general,
transitions for both K- and L- shell ions can be calculated with available
atomic codes. Nevertheless, in the past, the Fe L-shell lines have been
considered highly uncertain, presumably because of the increasing complexity
of multi-electron atoms. 

To address these issues, an x-ray spectroscopy laboratory-astrophysics program
was initiated 11 years ago and is still active today (PI's Kahn and
Beiersdorfer). 
The program is built around the unique capabilities of the \llnl\ \ebit\ 
to measure electron ion interactions. 
Driven by the astrophysical motivation, these
efforts have, so far, focused primarily on the Fe L-shell complex. In
particular, high-precision measurements for wavelengths and collisional 
excitation cross-sections (relative and absolute) have been published. 
Also, peculiar line ratios of Fe$^{16+}$ that have been puzzling
astronomers for years have been investigated and much of the atomic
uncertainty has been disentangled from the real astrophysical issues. 
For more details see the contribution of Brown et al. in these proceedings.

As the first astrophysical grating spectra became available, our team has made
a systematic attempt to test the ability of existing models to reproduce the
observed emission line intensities. 
It was found \citep {behar01a, brinkman01} that, generally, the observed line
intensities could be fairly well reproduced by state-of-the-art distorted wave
calculations. For instance, the \hullac\ code \citep {bs01} was used in those 
works. 
In particular, the Fe-L line intensities in the model including all of the
high-$n$ lines were found to fare quite well, implying that the calculated 
excitation rates and ensuing radiative cascades are fairly adequate. 
Conversely, accurate wavelengths still needed to be incorporated from
laboratory measurements. 
A more
detailed confrontation of the atomic calculations with Fe-L spectra of stellar
coronae can be found in \citet {behar01b}, where a comparison of calculations 
with the spectra of bright stellar coronal sources such as Capella
and HR 1099 confirms that, just like the K-shell atomic data, the Fe L-shell
data, when calculated correctly, are highly reliable and therefore very
useful. The latest versions of the widely used databases MEKAL \citep {mewe95}
and APEC \citep {smith01} now incorporate similar \hullac\ data
calculated by D. Liedahl with an earlier version of the code.

\section{Remaining Atomic Data Issues}

Although in general state-of-the-art models are doing well, several discrete,
but nonetheless important, discrepancies still remain. In particular, the
ratios of the 3s - 2p line intensities relative to those of the 3d - 2p
transitions of the same charge state were found to be anomalously high for
both Fe$^{16+}$ and Fe$^{17+}$. 
Similar effects have also been found in other late-type stars, and even in
elliptical galaxies \citep {xu02}, which suggests that they are not
associated with the astrophysical conditions. The origin of this discrepancy
has been recently studied in many theoretical and experimental works \citep
{laming01, doron02, peter02}, not all in agreement with each other.
Additionally, the L-shells of other cosmically abundant elements remain
largely unexplored. Major efforts in this direction are being conducted by
Lepson et al. (see these proceedings).
 
This is the place to note that the atomic processes associated with the x-ray
emission depend on the type of plasma at the source.
So far, we have focused on hot, collisional plasmas that are
governed by electron impact ionization and excitation. 
This type of plasma is relevant
to x-ray observations of stellar coronae, supernova remnants, the hot ISM in
old galaxies, the intergalactic haloes of galaxy clusters, and to the
overwhelming majority of laboratory plasmas. 
On the other hand, x-ray sources that are ionized and
excited by an external radiation field, e.g., active galactic nuclei 
(\agn) and x-ray binaries, require atomic data for photon impact processes.
Modeling line emission from photoionized
sources involves radiative recombination and photoexcitation rates
(i.e., oscillator strengths). In a recent paper on the x-ray spectrum of the 
type~II \agn\ NGC~1068 \citep {ali02}, we have shown that the
available data for the K-shell ions are very good for reproducing the x-ray
spectra from photoionized sources. Work is in progress to test the
status of the Fe-L data for photoionized sources. 
Similar data are routinely used to model absorption spectra. 
In absorption, however, one also observes inner-shell transitions, for which
 until recently there was an enormous lack of atomic data.

\section{The Urgent Need for Inner-shell Absorption Measurements}

The grating spectrometers on board \chandra\ and \xmm\ enable us for the first
time to detect x-ray absorption lines due to inner-shell photoexcitation. 
Since gas
under almost any conditions absorbs x-rays, these lines are ubiquitous to
x-ray absorption observations. Detections of inner-shell absorption lines have
been reported mostly in the ionized outflows of \agn, but also for absorption
by neutral ISM. 
Inner-shell absorption can probe the entire range of
ionization states from neutral up to highly-ionized Li-like species.
Consequently, these lines impose unprecedented, strict, constraints on the
ionization structure in the absorbing medium. 
Since there were extremely few relevant atomic data for these features in the
literature, in order to analyze the spectra, we had to embark on a tedious
endeavor of calculating numerous lines \citep {behar01c, behar02}. 
Calculations by the Ohio State team have also contributed to this
effort \citep {pradhan00, nahar01, pradhan02}. None of these
calculations have been benchmarked in the laboratory.

One case where x-ray absorption lines (including inner-shell excited) are 
particularly useful is for measuring velocities in AGN outflows \citep
[e.g.,][]{kaspi01}.
For these measurements, the rest frame wavelengths of the lines
need to be known to very high accuracies. The case of inner-shell K$\alpha$
absorption by oxygen is particularly interesting because it could potentially
relate the traditional x-ray absorber (O$^{6+}$ and O$^{7+}$) with absorbers
of other wavebands (e.g., O$^{5+}$ in the UV). 
Whether these absorbers represent the same kinematic systems or not has been
debated in the \agn\ community for some time now.
The correct diagnostics of the inner-shell
absorption lines could potentially provide a conclusive answer to this
interesting astrophysical question. In order to demonstrate the large
uncertainties of the currently available atomic data, in Table 1 we present
three different calculations for the wavelengths of the strongest K$\alpha$
inner-shell absorption lines of O$^{1+}$ through O$^{5+}$ and also compare
them with the deduced wavelengths from the \chandra/\letg\ observation of 
NGC~5548, courtesy of J. Kaastra. The wavelengths from NGC~5548 were obtained
assuming that the outflow velocities of all ions are similar to that of the
well-calibrated O$^{6+}$ velocity. 
Actually, this assumption may not be valid, but with the lack of laboratory measurements, it provides a rough idea of where to expect these lines.

Although one might have expected the R-matrix method to be the most rigorous,
it is clear from the table that at the current state of the atomic data, it is
virtually impossible to determine what are the correct rest frame positions of
these lines. The discrepancies among the various methods reach 50 m\AA, which
corresponds here to uncertainties in the measured velocities of 700 km/s. This
uncertainty is of the same order as typical outflow velocities in nearby
active galaxies, implying that these lines are practically useless for this
purpose, despite their great potential. Laboratory measurements are
desperately needed.

\section{Suggested Z Pinch Measurements}

The currently best available facility for producing inner-shell absorption
lines and measuring their wavelengths and optical depths is the z pinch at
Sandia National Laboratory. The powerful z pinch experiments (x-ray fluxes
reaching 10$^{19}$ erg s$^{-1}$ cm$^{-2}$) produce long-lived (6 ns), steady-
state, photoionized gas \citep {bailey01}, which can be studied with high
resolution spectrometers, both in emission and in absorption
simultaneously. The ionizing spectrum can be characterized rather accurately
by a blackbody spectrum.
The control over the position and density of the targets in these 
experiments provides a sensitive handle on the ionization state and column 
density in the absorbing medium.
A demonstration of these capabilities was
given by \citet {bailey01}, where absorption by photoionized Ne has been
measured. The spectrum obtained with a crystal spectrometer in that
experiment shows many individual lines that are nicely
resolved, which allows for accurate wavelength and equivalent-width
measurements.

\section{Summary and Prospects}

Many atomic data needs for hot, collisional plasmas relevant to x-ray
spectroscopic observations, now regularly obtained with the gratings on board
\chandra\ and \xmm, have been provided by the ongoing \llnl\ \ebit\ programs of
Kahn and Beiersdorfer. Particularly, these programs have provided the most
important data for Fe-L emission by hot gas and work is in progress to measure
many more high-quality data for other L-shell systems. Now that we actually
have high-resolution cosmic spectra, we can determine better than before which
atomic data are the most crucial. Thus, the most urgent needs of the x-ray
astronomy community for collisional plasmas will continue to be addressed 
with the \ebit\ Laboratory Astrophysics Program. 
Photoionized gases have received less
attention as they are more rare in nature and very few laboratory experiments
have sufficient x-ray flux to produce them. Particularly missing are
measurements for inner-shell absorption lines. The current uncertainties in
the positions of these lines considerably limit our ability to use them for
diagnostics. In the future, we intend to try to use z pinch experiments to
remedy this deficiency.

\clearpage




\begin{deluxetable}{lcccc}
\tablecaption{Wavelengths in \AA\ for the strongest 1s - 2p inner-shell
absorption lines in O ions. \label{tbl-1}}
\tablewidth{0pt}
\tablehead{
\colhead{Ion} & 
\colhead{R-matrix \tablenotemark{a}}   & 
\colhead{Cowan's Code \tablenotemark{b}}  & 
\colhead{\hullac\ \tablenotemark{c}} &
\colhead{NGC~5548 \tablenotemark{d}} 
}
\startdata

O$^{1+}$ & 23.27 & 23.31 & 23.30 & \nodata \\[0.1cm]

O$^{2+}$ & 23.08 & 23.10 & 23.11 & 23.17 $\pm$ 0.01 \\
         & 23.02 & 23.08 & 23.05 & 23.00 $\pm$ 0.02 \\
         & 22.93 & 23.01 & 22.98 & \nodata \\[0.1cm]

O$^{3+}$ & 22.73 & 22.77 & 22.73 & 22.74 $\pm$ 0.02 \\
         & 22.67 & 22.76 & 22.78 & \nodata \\
         & 22.67 & 22.76 & 22.73 & \nodata \\[0.1cm]

O$^{4+}$ & 22.35 & 22.38 & 22.33 & 22.38 $\pm$ 0.01 \\[0.1cm]

O$^{5+}$ & 22.05 & 22.05 & 22.00 & 22.01 $\pm$ 0.01 \\
         & 21.87 & 21.85 & 21.79 & \nodata \\

\enddata


\tablenotetext{a}{From \citet {pradhan02}}
\tablenotetext{b}{Raassen \& Kaastra, private communication}
\tablenotetext{c}{Present work}
\tablenotetext{d}{Kaastra, private communication}

\end{deluxetable}




\begin{thebibliography}{}

\bibitem[Bailey et al.(2001)]{bailey01} Bailey, J., et al. 2001,
J. Quant. Spectr. Radiat. Transfer, 71, 157

\bibitem[Bar-Shalom, Klapisch, \& Oreg(2001)]{bs01} Bar-Shalom, A., Klapisch,
M., \& Oreg, J. 2001, J. Quant. Spectr. Radiat. Transfer, 71, 169

\bibitem[Behar, Cottam, \& Kahn(2001a)]{behar01a} Behar, E., Cottam, J., \& Kahn, S.M. 2001a, \apj, 548, 966

\bibitem[Behar et al.(2001b)]{behar01b} Behar, E., Cottam, J., Peterson, J.,
Sako, M., Kahn, S.M., Bar-Shalom, A., Klapisch, M., \& Brinkman A.C. 2001b in
{\it X-Ray Astronomy 2000} ASP Conference Series Vol. 234, ed. R. Giaconni,
S. Serio, \& L. Stella (ASP, San Francisco), page 85

\bibitem[Behar, Sako, \& Kahn(2001c)]{behar01c} Behar, E., Sako, M., \& Kahn,
S.M. 2001c, \apj, 563, 497

\bibitem[Behar \& Netzer(2002)]{behar02} Behar, E. \& Netzer, H. 2002, \apj,
570, 165

\bibitem [Beiersdorfer et al.(2002)]{peter02} Beiersdorfer, P., et al. 2002,
\apj\ Letters, submitted

\bibitem[Brinkman et al.(2001)]{brinkman01} Brinkman, A.C., et al. 2001, \aap, 365, L324


\bibitem[Doron \& Behar(2002)]{doron02} Doron, R., \& Behar, E. 2002, \apj, 574
July 20 issue

\bibitem[Kaspi et al.(2001)]{kaspi01} Kaspi, S., et al. 2001, \apj, 554, 216

\bibitem[Kikhabwala et al.(2002)]{ali02} Kinkhabwala, A., et al. 2002, \apj, in
press (astro-ph/0203290)

\bibitem[Laming et al.(2001)]{laming01} Laming, J.M., et al. 2000, \apj, 545, L161

\bibitem [Mewe, Kaastra, \& Liedahl(1995)]{mewe95} Mewe, R., Kaastra, J.S., \& Liedahl, D.A. 1995, Legacy 6, 16

\bibitem[Nahar, Pradhan, \& Zhang (2001)]{nahar01} Nahar, S.N., Pradhan, A.K.,
\& Zhang, H.L. 2001 \pra, 63, 060701(R) 

\bibitem[Peterson et al.(2001)]{peterson01} Peterson, J.R., et al. 2001, \aap, 365, L104

\bibitem[Pradhan(2000)]{pradhan00} Pradhan, A.K. 2000 \apj, 545, L165

\bibitem[Pradhan et al.(2002)]{pradhan02} Pradhan, A.K., Chen, G.X., Delahaye,
F, Nahar, S.N., \& Oelgoetz, J. 2002, astro-ph/0204116

\bibitem[Smith et al.(2001)]{smith01} Smith, R.K., Brickhouse, N.S., Liedahl,
D.A., \& Raymond, J.C. 2001, \apj, 556, 69

\bibitem[van der Heyden et al.(2002)]{kurt02} van der Heyden, K., Behar, E.,
Vink, J., Rasmussen, A.P., Kaastra J.S., Bleeker, J.A.M., \& Kahn, S.M., \aap,
in press (astro-ph/0203160)

\bibitem[Xu et al.(2002)]{xu02} Xu, H., et al. 2002, \apj, submitted (astro-ph/0110013)

\end{thebibliography}
\end{document}